\documentclass[rsi, amsmath,amssymb,reprint, onecolumn]{revtex4-1}

\usepackage{graphicx}
\usepackage{dcolumn}
\usepackage{bm}

\usepackage[utf8]{inputenc}
\usepackage[T1]{fontenc}
\usepackage{mathptmx}
\usepackage{etoolbox}

\usepackage{siunitx}
\usepackage{subcaption}

\makeatletter
\def\@email#1#2{%
 \endgroup
 \patchcmd{\titleblock@produce}
  {\frontmatter@RRAPformat}
  {\frontmatter@RRAPformat{\produce@RRAP{*#1\href{mailto:#2}{#2}}}\frontmatter@RRAPformat}
  {}{}
}%
\makeatother
\begin{document}

\preprint{AIP/123-QED}

\title[]{Design and First Tests of the Trapped Electrons \\ Experiment T-REX}
\author{F.~Romano*}
\email{francesco.romano@epfl.ch}
\author{G.~Le~Bars}
\author{J.~Loizu}
\author{M.~Nöel}
\author{J.-P.~Hogge}
\author{S.~Alberti}
\author{J.~Genoud}
\author{S.~Antonioni}
\author{L.~Naux}
\author{P.~Giroud-Garampon}
\author{S.~Couturier}
\author{T.~Leresche}
\author{D.~Fasel}
 \affiliation{\'Ecole Polytechnique Fédérale de Lausanne (EPFL), Swiss Plasma Center (SPC), CH-1015 Lausanne, Switzerland.}

\date{\today}

\begin{abstract}

Gyrotrons are essential for electron cyclotron resonance heating (ECRH) in fusion reactors, making efficient operation crucial for advancing fusion energy. Past experiments revealed instability issues due to trapped electrons in the magnetron injection gun (MIG) region, causing undesired currents and operational failures. To address this, tight manufacturing tolerances are required for the MIG geometry~\cite{pago2}. We present initial findings of the TRapped Electrons eXperiment (T-REX) developed at the Swiss Plasma Center, designed to understand the physics of electron clouds in gyrotron MIGs. T-REX replicates MIG geometries, as well as their typical electric and magnetic fields, and it is supported by 2D Particle-in-Cell (PIC) simulations with the FENNECS code~\cite{guilPoP,guilth}. The setup includes two coaxial electrodes in a vacuum chamber atop a superconducting magnet, with a central electrode biased to negative DC voltages and an outer one at ground, creating a radial electric field ($1$ to $\SI{2}{\mega\volt\per{\meter}}$) and an axial magnetic field ($B < \SI{0.4}{\tesla}$). This setup mimics Penning-Malmberg traps. We present the experimental device and first findings on current distribution and also qualitative comparison with FENNECS simulations~\cite{FENNECS}. Planned diagnostics include optical emission spectroscopy, phosphor screen imaging, Streak camera imaging, and potentially electric field distribution via the Stark effect. This research aims to enhance gyrotron performance and reliability in fusion energy systems.

\end{abstract}

\maketitle

\section{\label{sec:introduction}Introduction}

\begin{figure}[h]
    \centering
    \includegraphics[width=.5\linewidth]{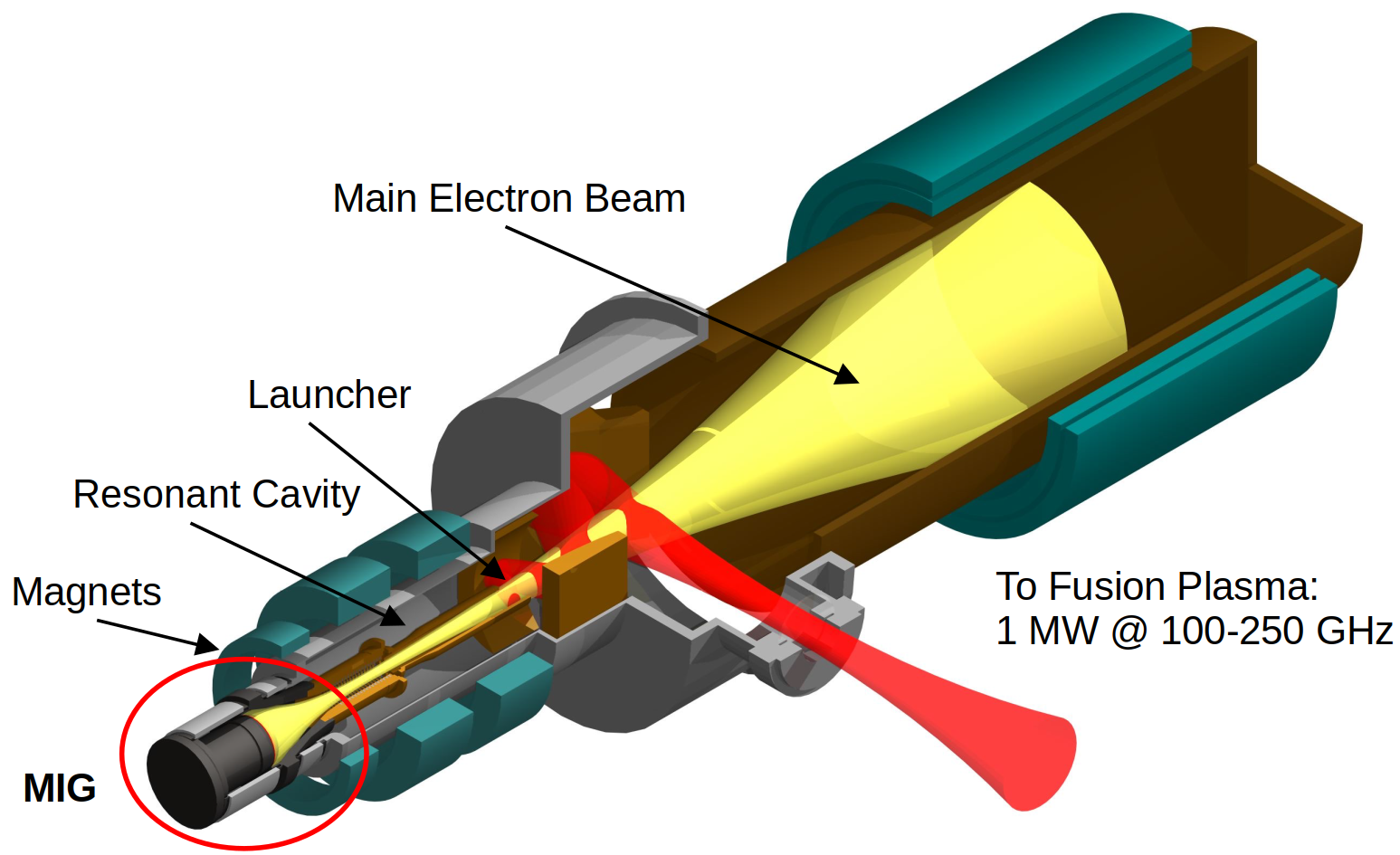}
   \caption{Schematics of a gyrotron and its main components. The red ellipse represents the gyrotron region where the secondary electrons (i.e. not belonging to the main electron beam) can potentially form a trapped electron cloud which is the object of this article. Adapted with permission from~\cite{alberti}: S. Alberti, J. Genoud, T. Goodman, J.-P. Hogge, L. Porte, M. Silva, T.-M. Tran, M.-Q. Tran, K. Avramidis, I. Pagonakis, et al., in EPJ
Web Conf., Vol. 157 (2017) p. 03001.}
    \label{fig:gyrotron}
\end{figure}

To achieve future fusion energy, both Tokamak and Stellarator configurations foresee the use of gyrotrons as the main plasma heating devices via electron cyclotron resonance heating (ECRH) and/or current drive (ECCD)~\cite{stellarator,ITER}. The fusion reactor ITER currently requires 24 \SI{}{\mega\watt}-level gyrotrons~\cite{10157606,10157774}, operating at \SI{170}{\giga\hertz}, possibly increasing to 48 or even 60 after recent changes to the wall material~\cite{Gruber_2009,TERENTYEV2024114200}. For DEMO, instead, the current plan is to deliver \SI{130}{\mega\watt} of power via gyrotrons~\cite{DEMO}. Therefore, their efficient operation is critical, and any design modification that can ease construction requirements and tolerances will reduce costs and manufacturing time~\cite{pago1}. Gyrotrons are high power devices that produce an electromagnetic wave at frequencies in the sub-\SI{}{\tera\hertz} range. In a gyrotron, see Fig.~\ref{fig:gyrotron}, electrons are generated via thermionic emission by the magnetron injection gun (MIG). The electrons are accelerated at mildly relativistic energies in a crossed configuration of electric and magnetic fields. They rotate at the relativistic cyclotron frequency, which depends mainly on the applied magnetic field amplitude, but also weakly on their energy. This defines the gyrotron operating frequency. The annular electron beam travels inside a cylindrical resonant cavity in which a specific electromagnetic wave mode is excited. The resonant interaction between the gyrating electrons and the electromagnetic mode supported by the cavity leads to a net transfer of energy from the electrons to the wave. After exiting the cavity, the excited transverse electric (TE) mode is converted into a Gaussian beam which is then propagated through an output window. They gyrotron very high-level vacuum is ensured by its window.

This article focuses on the development of a novel and unique non-neutral plasma physics experiment with the aim of understanding a specific issue found in some gyrotron MIGs. For specific conditions, nominal gyrotron operation could not be achieved and/or unexpected shutdown of the gyrotron happened~\cite{pago2}. The cause has been tracked down to secondary electrons trapped (those that do not belong to the main electron beam) in the MIG region due to the presence of potential wells. Those can form when magnetic field lines with a specific curvature cross twice equipotential lines. The potential well size and depth depend on the combination of the applied magnetic field and the geometry of the electrodes to which a potential is applied. Potential wells can trap electrons and, via electron leakage from the trap, can cause relatively large currents to arise, leading to possible internal gyrotron damages and preventing the power supply from sustaining the nominal, externally applied, voltage bias. This is, specifically, the body power supply (BPS) that determines the electron beam properties and is typically designed for \SI{40}{\kilo\volt} and \SI{150}{\milli\ampere}~\cite{gyroBPS}.
The current workaround is of designing the electrodes based on the vacuum magnetic and electric equipotential field lines to minimize the presence of potential wells. Gyrotrons are sealed vacuum tubes and it is difficult to implement diagnostics to study and characterize the dynamics of the trapped electron clouds, therefore, this calls for a dedicated basic plasma experiment to recreate such clouds in a flexible environment and with dedicated diagnostics. In fact, a deeper understanding of the clouds dynamics can lead to the establishment of new solutions that might simplify future gyrotron construction and manufacturing.

In this article, we present the TRapped Electrons eXperiment (T-REX), designed and built at the Swiss Plasma Center of the EPFL. Its primary objective is to explore the formation and evolution of electron clouds in MIG designs.
The article starts with the basics of non-neutral plasma physics and a brief overview on electron trapping mechanisms. Next, the T-REX facility is presented in detail and early operation results are shown and discussed. Finally, a preliminary comparison between T-REX experimental results and simulation results obtained with the code FENNECS, developed at the Swiss Plasma Center, is given. This is then completed by the plan for future diagnostics.  

\section{\label{sec:electrontrapping} Non-neutral Plasma and Electron Trapping}

\begin{figure}
    \centering
    \includegraphics[width=.3\linewidth]{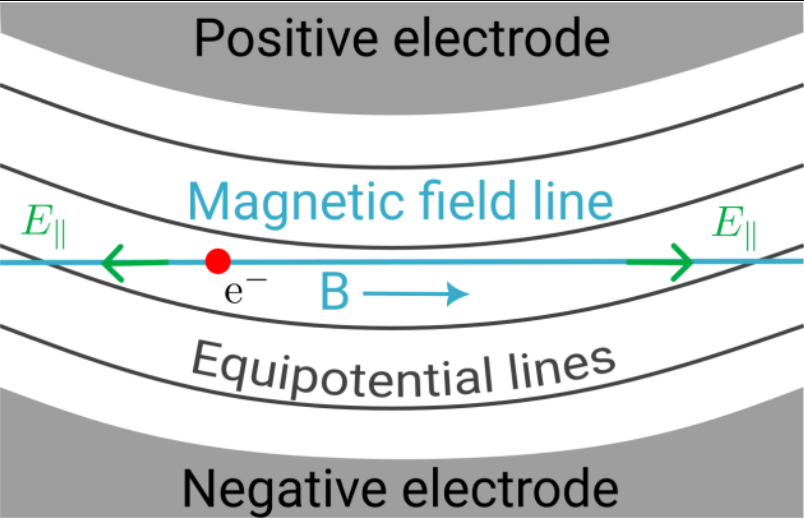}
\caption{Electron trapping via electric and magnetic fields: a magnetic field line is shown crossing twice an equipotential field line with an electron trapped. Adapted with permission from G. M. Le Bars, Modelling of nonneutral plasmas trapped by electric and magnetic fields relevant to gyrotron electron guns, Ph.D. thesis,
EPFL, Lausanne (2023)~\cite{guilth}.}
    \label{fig:trapping}
\end{figure}

The study of electron clouds (and electron trapping) lies within the field of non-neutral plasmas, which are plasmas without overall charge neutrality and dominated by one species. These plasmas have been studied since the 1960s~\cite{Davidson1991PhysicsON}. Electrons can be confined in various ways, primarily through magnetic mirrors or a combination of electric and magnetic fields, see Fig.~\ref{fig:trapping}.

\begin{figure}
    \centering
    \includegraphics[width=.6\linewidth]{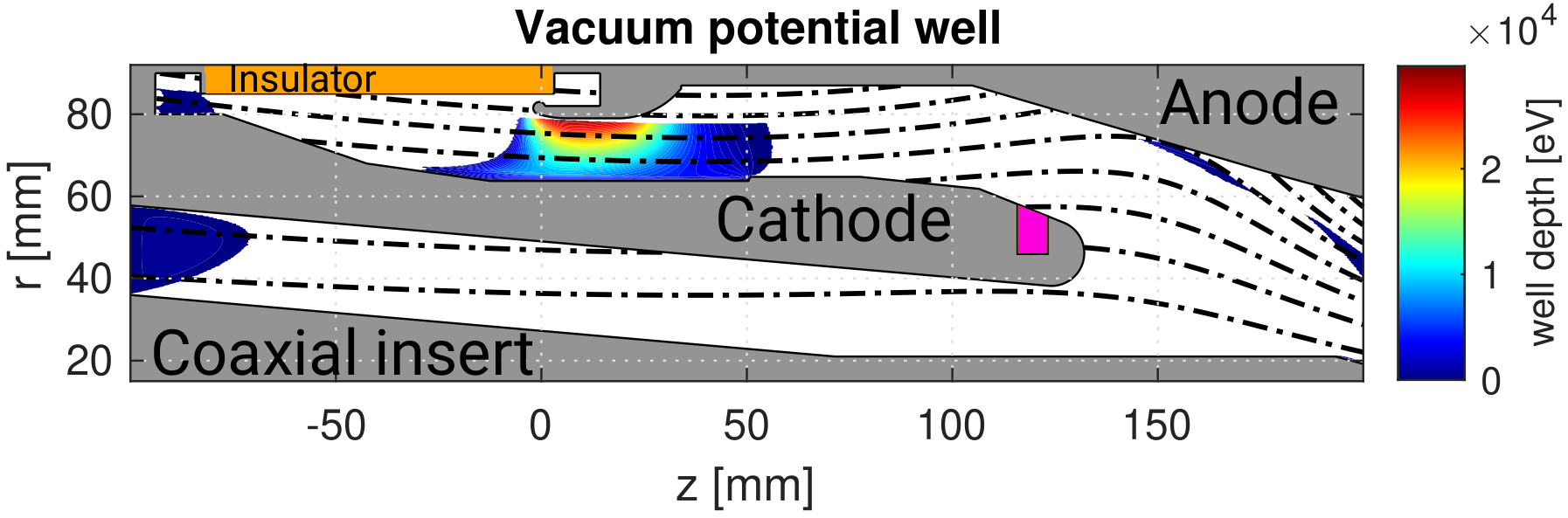}
\caption{Vacuum potential well in the MIG of the first European \SI{170}{\giga\hertz} - \SI{2}{\mega\watt} coaxial gyrotron prototype developed for ITER~\cite{pago3}. The cathode is at \SI{-90}{\kilo\volt} and the other electrodes are grounded. The black dashed-dotted lines are the magnetic field lines, and the magenta surface highlights the emitter ring, where the main electron beam is produced. The $z-$axis (at $r=0$) corresponds to the axis of azimuthal symmetry. The coloured areas in the vacuum region indicate different potential wells with the main one reaching a depth of \SI{30}{\kilo\electronvolt} as indicated by the color scale. Adapted with permission from G. M. Le Bars, Modelling of nonneutral plasmas trapped by electric and magnetic fields relevant to gyrotron electron guns, Ph.D. thesis,
EPFL, Lausanne (2023)~\cite{guilth}}
    \label{fig:MIG}
\end{figure}

Due to the electric and magnetic field topology of a gyrotron's MIG, the resulting trapping mechanism is the one that uses a combination of electric and magnetic fields. Consequently, it is also employed in T-REX. Specifically, such trapping system is resembling to the Penning-Malmberg trap~\cite{Malmberg1,Malmberg2,Malmberg3}. This trap consists of a stack of annular ring-shaped coaxial electrodes placed in a purely axial magnetic field. The electrodes are biased at different potentials to create an axially confining electric field. Penning-Malmberg traps are commonly used to trap and study elementary particles at very low temperatures. Similarly, a gyrotron's MIG, see Fig.~\ref{fig:MIG}, provides a nearly axial magnetic field, while the electric field is shaped by coaxial electrodes biased at different voltages.
In a gyrotron's MIG, the equipotential lines are shaped along with the MIG's electrodes geometry, and the magnetic field line geometry is determined by external magnets. The magnetic field is azimuthally symmetric and, while it has to be uniform in the resonant cavity of a gyrotron, it is non-uniform in the MIG region. The electric field, on the other hand, is mainly radial and imposed by the different electrodes. In the MIG region, vacuum potential wells can arise where the magnetic field lines cross twice an equipotential line that is shaped by the electrodes geometry, see Fig.~\ref{fig:trapping}~\cite{pago2}.

In MIGs, the trapped electrons are not those belonging to the main gyrotron electron beam. The thermoionically generated electrons are accelerated from the annular emitter towards the cavity. The crossed electric and magnetic fields allow to impart both perpendicular (to the magnetic field) and parallel kinetic energy. At the nominal conditions, all electrons forming the annular beam leave the emitter and eventually reach the gyrotron collector, see Fig.~\ref{fig:gyrotron}. The secondary electrons generation mechanisms are diverse and can involve several distinct sources. 
The mechanisms for (secondary) electron generation can be summarized as follows:
\begin{enumerate}
    \item Cosmic rays can free electrons by hitting neutrals;
    \item Field ionization: the large electric field between electrodes can induce ionization of the residual gas (not simulated in FENNECS); 
    \item Field emission: electrons emitted from the biased surfaces due to large electric fields;
    \item Ion induced electron emission: free electrons can be emitted from ion impact on the surfaces; 
    \item Electron impact ionization of neutrals: free electrons impacting neutrals create more free electrons.
\end{enumerate}

As the amount of charge increases in the potential well, the space charge of the electron cloud also modifies the potential well, and, finally it can create a path for undesired currents to flow via the electron cloud between the electrodes surfaces. Such build up of charge can lead to possible instabilities leading to detrimental leakage currents, causing issues to the gyrotron operation as well as damaging it. The current design criteria for gyrotron MIGs include the avoidance of such vacuum potential wells by shaping the electrodes according to the nominal magnetic field topology. These requirements are very challenging from the point of view of engineering and manufacturing, and also requires a time consuming iterative design process. Finally, this also reduces the gyrotron's operating space in terms of magnetic field, since the magnetic field line are frozen. 

\section{The TRapped Electrons eXperiment - T-REX}

\begin{figure}
    \centering
    \includegraphics[width=.8\linewidth]{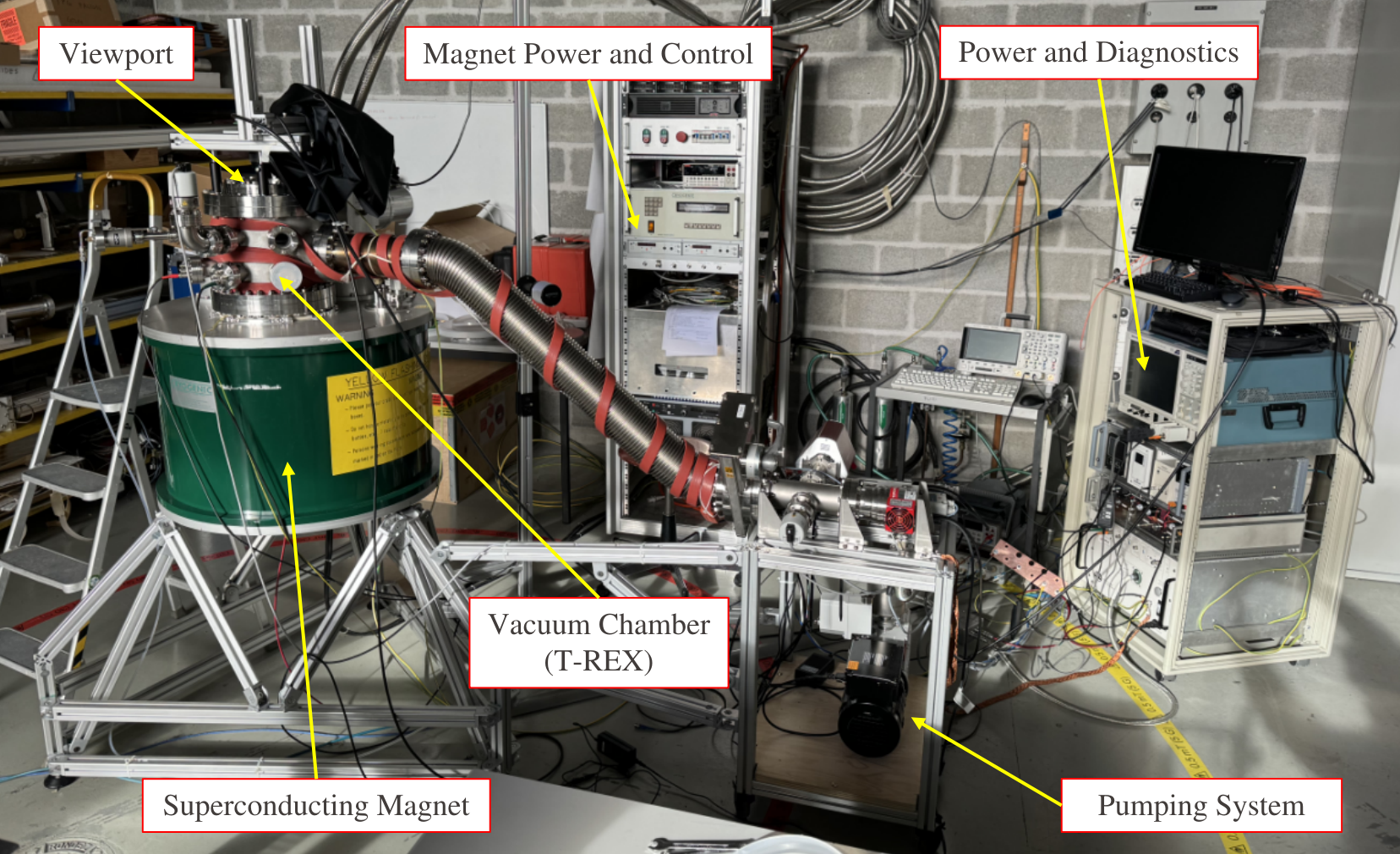}
   \caption{Picture of the T-REX facility. Left: superconducting magnet with the vacuum chamber containing T-REX on top. Behind: the magnet power and controls rack. Middle front: pumping system. Right: power and diagnostics rack.}
    \label{fig:T-REX_Photo}
\end{figure}

\begin{figure}
    \centering
    \includegraphics[width=0.55\linewidth]{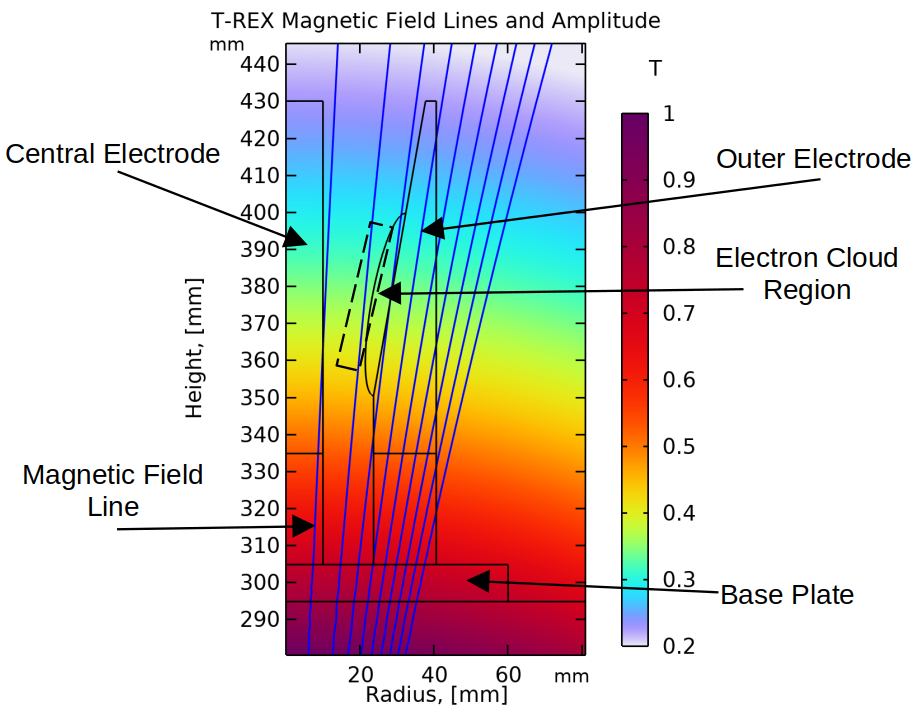}
    \caption{T-REX magnetic field lines and amplitude. The magnet is set at maximum field. The maximum magnet field amplitude in the cloud region is $B<\SI{0.4}{\tesla}$.}
    \label{fig:T_REX_Magnetic}
\end{figure}

\begin{figure}
    \centering
    \includegraphics[width=.83\linewidth]{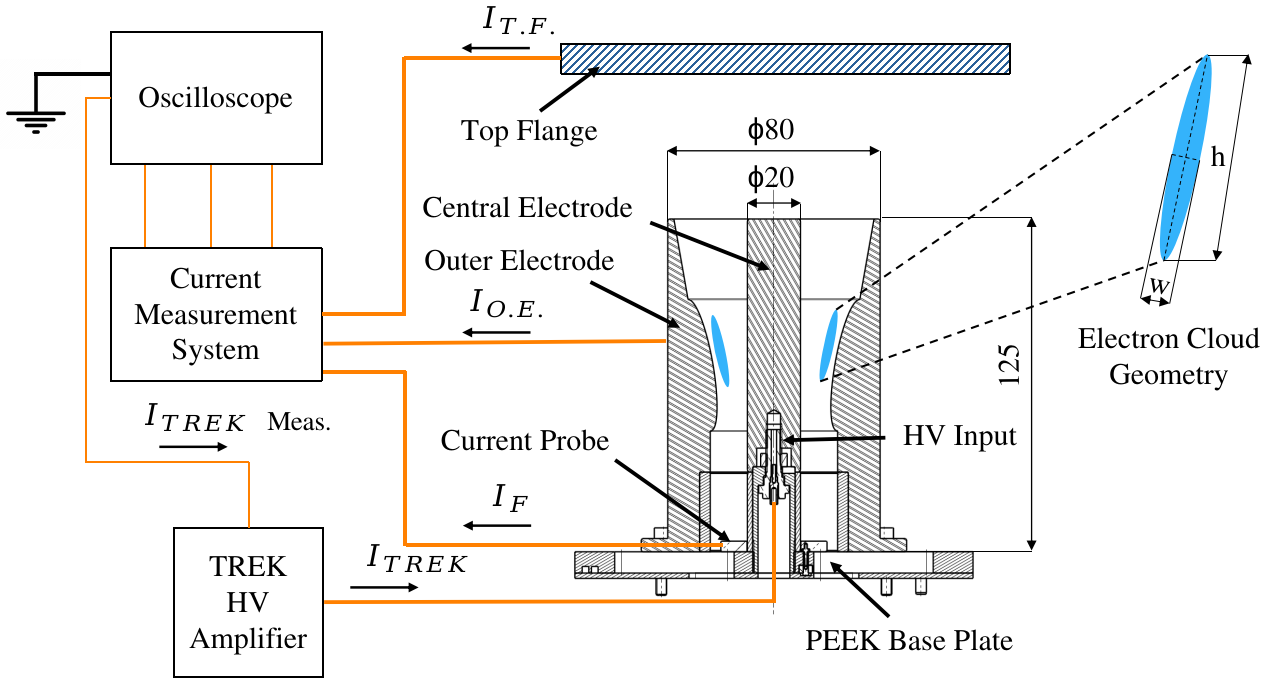}
    \caption{Sectioned view of T-REX electrodes, current measurement set-up, definition of electron cloud geometry, $w$ radial thickness, $h$ vertical height. Dimensions are in [\SI{}{\milli\meter}].}
    \label{fig:measure}
\end{figure}

T-REX is a basic plasma experiment aiming at studying electron cloud formation in gyrotron's MIG-like configurations without the presence of primary electrons. The concept is to recreate electron trapping conditions similar to those that are problematic in gyrotron's MIG, namely a potential well produced by the combination of electric and magnetic fields and the presence of trapped electrons in it. An image of the whole setup is shown in Fig.~\ref{fig:T-REX_Photo}.
The core of the experiment is composed of two coaxial electrodes whose shape is such to form a potential well mimicking the one existing in some MIG's: the central electrode is biased at a negative high-voltage (HV), while the outer electrode is at ground. Those are installed within a vacuum chamber placed on top of a superconducting magnet. The geometry of the electrodes shapes the vacuum electric field, while the magnetic field is slightly divergent in the trapping region, see Fig.~\ref{fig:T_REX_Magnetic}. Both, the electric and magnetic fields amplitude in the potential well are similar to the ones in a real MIG, however, the vacuum level in this same region is significantly larger than in the real MIG. This last point is due to the fact that the entire gyrotron is baked under vacuum at temperatures exceeding \SI{400}{\celsius}, conditions which are not possible to obtain in T-REX. The superconducting magnet is in a cryogen-free cryostat, where Helium remains in its gaseous phase in a closed loop system connected to a compressor. The magnet generates up to $B=\SI{9.7}{\tesla}$ at the centre of its borehole, whereas in the area of the electron cloud the maximum magnetic field is $B<\SI{0.4}{\tesla}$, see Fig.~\ref{fig:T_REX_Magnetic}.
The vacuum chamber is mounted on top of the superconducting magnet and is provided with multiple side access flanges to install electrical feedthroughs and/or viewports. To study the electron cloud behavior on multiple gases, gas injection is available. This can also be used to regulate the pressure inside the vacuum chamber $p$. The target vacuum level is of high vacuum, with a target pressure range between $p=\SI{1E-7}{}-\SI{1E-5}{mbar}$. In a gyrotron, the vacuum is expected to be significantly better, specifically in the range of \SI{1e-9}{\milli\bar} as it is a sealed baked system. The HV is delivered to T-REX via a TREK HV amplifier 20/20C that provides up to $\pm \SI{20}{\kilo\volt}$ DC or peak AC at up to $\pm \SI{20}{\milli\ampere}$ DC or peak AC. The TREK power supply is a fast amplifier covering a bandwidth frequency up to \SI{3.75}{\kilo\hertz} for large amplitude variations, or \SI{20}{\kilo\hertz} for 
a small one. The present results have been obtained operating the TREK in DC operation mode. A \SI{5}{\meter} HV cable connects the TREK HV output terminal to the bottom of the vacuum chamber via a vacuum feedthrough that is galvanically connected to the T-REX central electrode. The coaxial electrodes of T-REX are made of 6082 aluminium alloy and are interchangeable, meaning that multiple combinations of geometries can be tested. At the base of the coaxial electrodes, a copper ring is installed operating as a current probe to measure the amount of electrons that travels downwards following the magnetic field lines, see Fig.~\ref{fig:measure}. The electrodes assembly is mounted on top of a polyether ether ketone (PEEK) plate which electrically isolates it from the vacuum chamber, and also provides guiding for the wiring to the electrical feedthroughs of the vacuum chamber. The design of the first test geometry, also shown in Fig.~\ref{fig:measure}, has been supported by simulations performed with the FENNECS code. FENNECS is a 2D $(r,z)$ PIC code~\cite{guilPoP2,guilth} that solves the Boltzmann-Poisson equation for electron and ion distribution functions that has been successfully validated against experiments in gyrotron MIG. For T-REX, the simulations showed that, with the available combination of magnetic and electric field topology (therefore electrodes geometry), and background gas pressure and composition, the resulting potential well forming between the two electrodes can trap electrons and lead to an electron cloud with the following properties:
\begin{itemize}
    \item Electron kinetic energy $E_k = 0.1-\SI{3}{\kilo\electronvolt}$, $v_{E\times B, max}=\SI{1E6}{\meter\per\second}$
    \item Maximum electron cloud density range: $n_{e,max}=\SI{1E17}{}-\SI{8E17}{\meter^{-3}}$
    \item Electron cloud (radial) thickness: $w=\SI{1.5}{\milli\meter}$, vertical height: $h=\SI{45}{\milli\meter}$, see Fig.~\ref{fig:measure}.
    \item Brillouin ratio $2\omega_p^2/\omega_c^2 = 0.5-1$ 
\end{itemize}

Where $\omega_p = \sqrt{{n_e q_e^2} / {m_e \epsilon_0}}$ is the plasma frequency, and $\omega_c= q_e B / m_e$ is the electron cyclotron frequency, with $n_e, m_e, q_e$ the electron density, mass, and charge respectively, and $\epsilon_0$ the vacuum permittivity. The electron cloud geometry is depicted in Fig.~\ref{fig:measure}. The Brillouin ratio measures the relative strengths of the (defocusing) space-charge force and the (focusing) magnetic force on the plasma~\cite{Davidson1991PhysicsON}, and in this case indicates that the electron cloud obtained in T-REX has a high density compared to Penning traps that usually have Brillouin ratios much smaller than $<1$~\cite{guilth}.

Finally, the experiment control parameters are:
\begin{itemize}
    \item Magnetic Field Amplitude $B$; 
    \item Electric Field Amplitude via $V_{bias}$;
    \item Vacuum Level $p$;
    \item Gas Composition;
    \item Electrodes Geometry (Electric Field Shape).
\end{itemize}

For the first test campaign, current and voltage are measured. In particular, the input and the resulting currents flowing to the ground from the experiment main elements, such as the outer electrode, the current probe placed at the base of the electrode assembly, and the top flange of the vacuum chamber. The knowledge of such currents can provide a picture of the current/power distribution among the experiment main elements. A voltage probe is also installed on the electrical path between the HV power supply and T-REX to provide an additional reading on the output voltage.The current measurement system employs amplifiers circuits with an accuracy of $\pm\SI{0.05}{\milli\ampere}$ and a cut-off frequency of $<\SI{160}{\kilo\hertz}$. Each experiment main component is electrically insulated and grounded via single cables that are connected to an oscilloscope via the current measurement system. The experiment setup for this test campaign is shown in Fig.~\ref{fig:measure}.

\begin{figure}
    \centering
    \includegraphics[width=.75\linewidth]{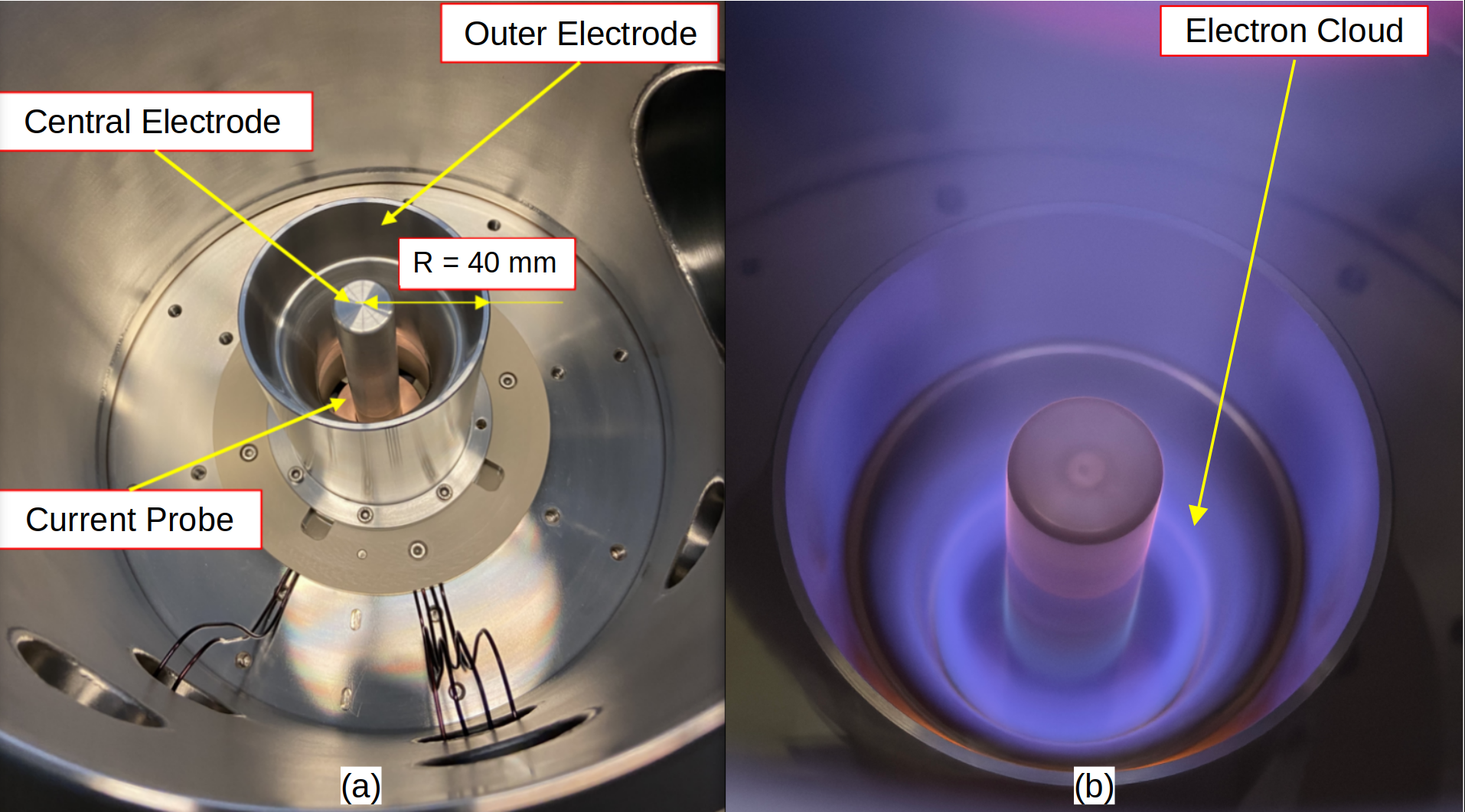}
    \caption{Inside view of T-REX: \textit{(a)} two coaxial electrodes, current probe, PEEK insulating base plate and cabling. \textit{(b)} T-REX in operation $B=\SI{0.31}{\tesla}$, $V_{bias}=\SI{-10}{\kilo\volt}$, He at $p=\SI{3E-4}{\hecto\pascal}$ highlighting the glowing electron cloud.}
    \label{fig:T-REXin}
\end{figure}

For the test campaign hereby described, the top of the experiment is closed via a blind stainless steel flange. Photos showing the coaxial electrodes assembly inside of T-REX vacuum chamber, with and without plasma, are presented in Fig.~\ref{fig:T-REXin}. Figure~\ref{fig:T-REXin}(a) shows the assembly of the electrodes, including the central and outer electrodes, the current probe at the bottom, and the cabling connecting to the electrical vacuum feedthroughs. In Figure~\ref{fig:T-REXin}(b), the T-REX device is depicted during operation. A ring-shaped electron cloud is visible near the outer electrode, in agreement with simulation predictions. The plasma light emission arises from excitation and de-excitation of Helium (in this case). An estimate based on cross-sections for different processes (ionization, recombination, excitation) and the very rapid loss of ions to the central electrode rule out recombination as a significant contributor to the light emission~\cite{guilth}. A qualitative calculation of excitation vs recombination rates is shown in Appendix~A.

From the results of FENNECS simulations, the formation and dynamics of the cloud can be understood as follows: a single free electron entering the potential well, originating from some of the aforementioned sources such as cosmic rays, field emission, or a finite ionization rate at room temperature, can trigger the exponential electron density growth. This initial electron quickly acquires $\vec{E} \times \vec{B}$ drift velocity in the azimuthal direction and initiates ionization of the background neutral gas through impact collisions. The resulting electrons, trapped in the well, further ionize the neutral gas, while the ions, due to their large Larmor radius, are lost upon striking the central electrode, potentially releasing additional electrons. This process leads to the spontaneous formation of a high-density electron cloud in the potential well. Recombination is negligible as there are no ions present. Collisions between cloud electrons and neutral gas generate a net azimuthal drag force, which, in combination with the axial magnetic field, induces radial drift towards the outer electrode. This drift results in electron loss at the outer electrode, though some electrons may escape earlier if their axial velocity surpasses the potential well.

A steady-state can be achieved when the cloud density reaches a point where the impact ionization rate is balanced by axial and radial drift losses. Consequently, currents are expected on the outer electrode, due to collisional radial drift losses, and on the top flange due to axial losses from the cloud and from the central electrode, where electrons produced by impacting ions are not trapped.

\section{T-REX Testing and First Results}
In September 2023, the first plasma was achieved with T-REX, followed by an initial assessment of the facility behavior based on pressure, starting voltage, magnetic field, and gas type. It has been observed that plasma is formed spontaneously as the applied voltage to the central electrode is above a certain threshold, namely the starting voltage. Plasma only forms if there is a magnetic field applied, the minimum required magnetic field is still object to study, but preliminary experiments set it to $\SI{0.05}{\tesla}<B<\SI{0.10}{\tesla}$ in the cloud region, its amplitude does not seem to change the starting voltage. Observation reveals an interaction between the plasma and the dielectric viewport composed of Kodial/7056 alkali borosilicate glass, which is mounted atop the vacuum chamber. As the viewport is not conductive, charged particles accumulate on it, likely modifying the plasma conditions and causing erratic behavior in the applied HV signal. In a second configuration, which will be the configuration used for the described test campaign, a blank stainless steel flange has been installed instead, electrically isolated from the vacuum chamber, and grounded. On one hand, this configuration provides a more controlled experiment environment even tough it limits the optical accessibility to the electron cloud region. On the other hand, it corresponds to the potential configuration that can be simulated with FENNECS. The use of viewports with an indium tin oxide (ITO) electrically conductive transparent coating is foreseen for the future. To characterize the behavior of T-REX, the objective of the first set of tests is to evaluate how the total input current is distributed between the main experiment components. Helium is set as input gas. 

An electron emitter has been initially foreseen for T-REX but finally, from experimental evidence, it has been found unnecessary. Plasma could be achieved without an extra source of electrons. Moreover, according to FENNECS simulations, an extra source of electrons, as long as the seed is much smaller than the amount of electrons that the chain reaction would produce, would not change the electron cloud parameters such as $n_{e,max}$ and size. With the magnetic field active, and above a minimum level, once a sufficiently large DC bias $V_{bias}$ is applied, the electron cloud forms between the two electrodes and can be visually seen, and the resulting currents measured. 

Five magnet settings have been selected for testing, corresponding to $B\sim 0.05, 0.10, 0.17, 0.24, \SI{0.31}{\tesla}$ in the expected electron cloud region. The measured background pressure with a constant Helium flow applied, is of $p=\SI{3.2E-4}{\milli\bar}$, measured on the side of the vacuum chamber via a capacitive pressure gauge. While the pressure is not measured directly where the cloud forms, our first objective is to measure the order of magnitude of the vacuum level. Multiple bias voltages have been tested, from first plasma voltage, down to $V_{bias}=\SI{-12}{\kilo\volt}$. Lower voltages are possible but not included in this first analysis. The current output from the TREK HV amplifier, $I_{TREK}$, is measured via the TREK itself and it is acquired using the oscilloscope. 

The currents flowing to the ground from the outer electrode, the current probe, and the top blank flange are named $I_{O.E.}, I_{F}, I_{T.F.}$ respectively, and measured via the aforementioned in-house current measurement system. Current measurements are performed with the experiment in DC operation, the length of the acquired signal is of \SI{0.2}{\micro\second} and the values presented here are time-averages.

T-REX is set to one input gas flow of Helium for a pressure in the range of $p=\SI{3E-4}{\milli\bar}$ and one magnetic field setting. First plasma is achieved, at $V_{bias}$ between $\SI{-1.2}{\kilo\volt}$ and $\SI{-1.5}{\kilo\volt}$. The bias for the first measurement point is applied directly at $V_{bias}=\SI{-2}{\kilo\volt}$ to make sure the electron cloud is present, then it is increased down to $V_{bias}=\SI{-12}{\kilo\volt}$ in steps of $\SI{1}{\kilo\volt}$. At each point, $I_{TREK}, I_{O.E.}, I_{F}, I_{T.F.}$ are measured. It is found that, for all cases, most of the current is collected on the top flange. 

The reason for this could be due to multiple sources such as:
\begin{itemize}
\item electrons escaping the potential well and travelling along the magnetic field lines upwards to the top of the vacuum chamber (already modeled in FENNECS);
\item glow discharge between the top of the central electrode and the top flange of the vacuum chamber (no model in FENNECS yet);
\item ion induced electron emission (IIEE) on the central electrode (already modeled in FENNECS).
\end{itemize}

\begin{figure}
\centering
    \begin{subfigure}{0.42\textwidth}
    \centering
    \includegraphics[width=\linewidth]{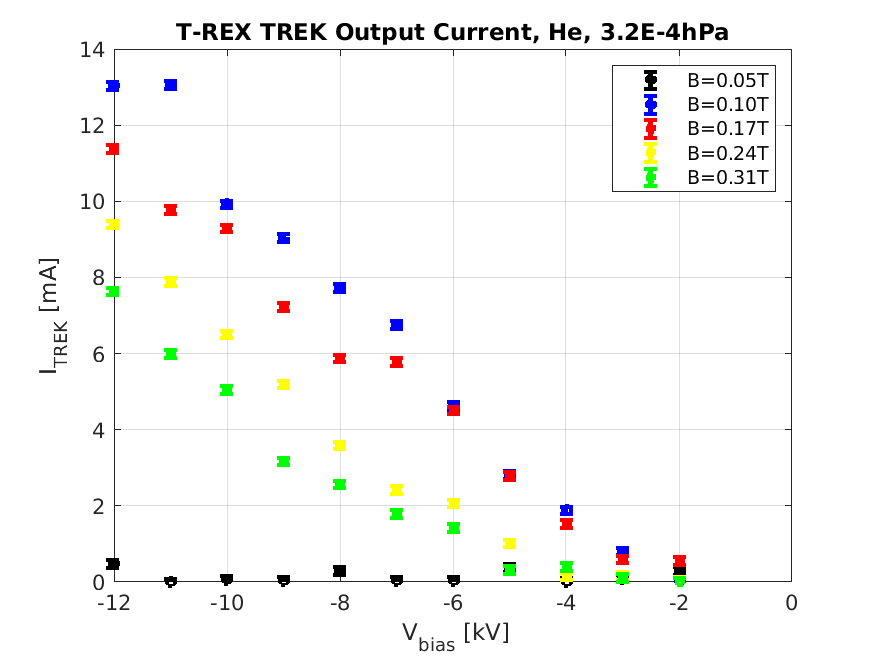}
    \caption{}
    \label{fig:ITREK}
    \end{subfigure}
    \begin{subfigure}{0.42\textwidth}
    \centering
    \includegraphics[width=\linewidth]{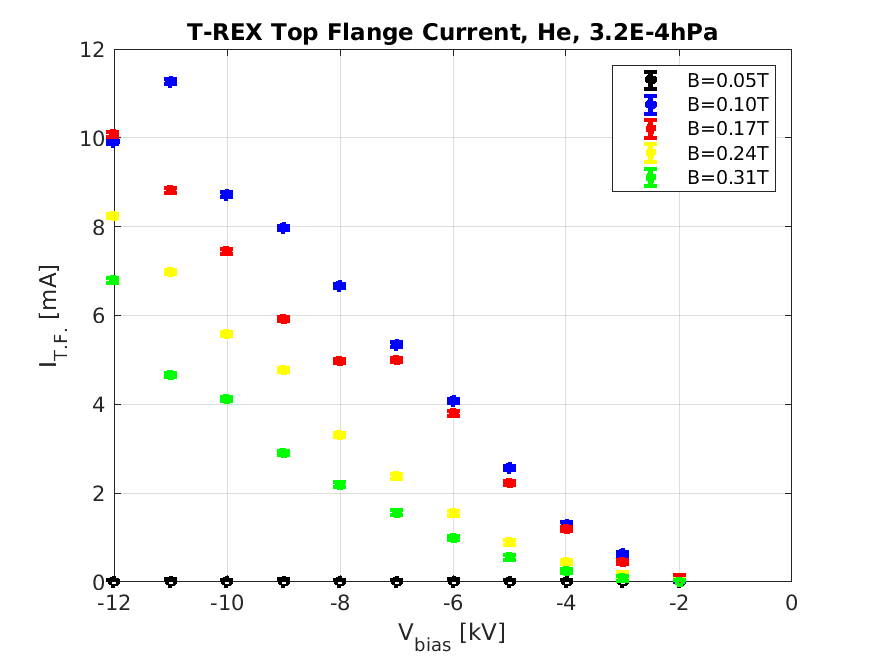}
    \caption{}
    \label{fig:ITF}
    \end{subfigure}
    \begin{subfigure}{0.42\textwidth}
    \centering
    \includegraphics[width=\linewidth]{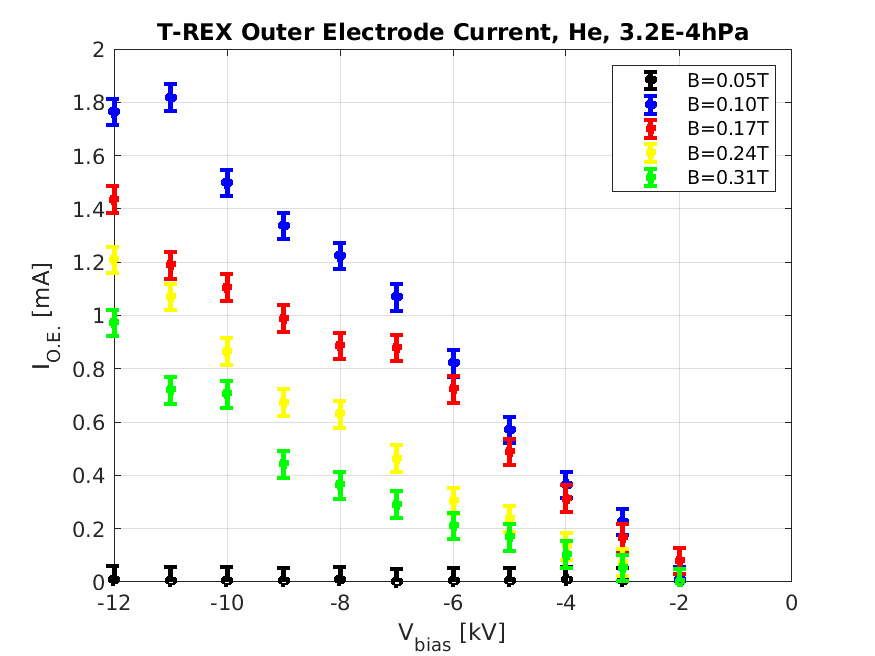}
    \caption{}
    \label{fig:IOE}
    \end{subfigure}
    \begin{subfigure}{0.42\textwidth}
   \centering
   \includegraphics[width=\linewidth]{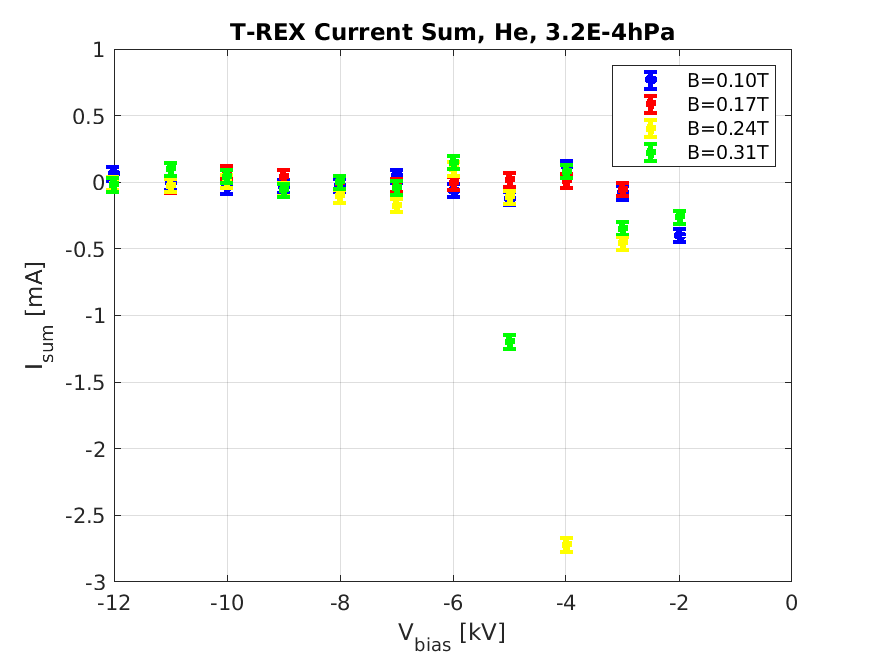}
   \caption{}
   \label{fig:ISUM}
   \end{subfigure}
   \caption{T-REX current measurements vs $V_{bias}$ and $B$, He at $p=\SI{3.2E-4}{\hecto\pascal}$, averaged values. (a) TREK Output Current $I_{TREK}$. (b) Current from Top Flange to Ground $I_{T.F.}$. (c) Current from Outer Electrode to Ground $I_{O.E.}$. (d) Sum of the Currents $I_{Sum}$.}
   \label{fig:Currents}
\end{figure}

The measured currents presented hereby are the DC components of the signal and allow the study of the current distribution. In particular, $I_{TREK}$, $I_{T.F.}$, and $I_{O.E.}$ are shown in Figs.~\ref{fig:ITREK}, \ref{fig:ITF}, and \ref{fig:IOE} as a function of $V_{bias}$, and for different $B$. The diagram of $I_F$ is not shown here as all points show zero current. The current $I_{Sum} = I_{TREK} - I_{T.F.} - I_{O.E.} - I_F$ is plotted in Fig.~\ref{fig:ISUM} as well, its value shall be $I_{Sum}\sim $ zero due to charge conservation, meaning that the measurement error is sufficiently small to allow the reconstruction of the current distribution in T-REX. Hereby, the error bars display the accuracy for the current measurement system at $\pm \SI{0.05}{\milli\ampere}$, while the accuracy of $I_{TREK}$ is $<\SI{0.2}{\milli\ampere}$. The condition at \SI{-11}{\kilo\volt} for the $B=\SI{0.10}{\tesla}$ case, led to an outlier current value, as one of the amplifiers broke during that measurement. This, however, does not change the general trend found in experiments for all the conditions. 

$I_{TREK}$ grows along with $V_{bias}$ up to $I_{TREK}<\SI{13}{\milli\ampere}$, and its amplitude is larger when decreasing the magnetic field, while for $B\leq \SI{0.05}{\tesla}$, there is no current as the cloud does not form. Also, with the absence of magnetic field, we were not able to generate any plasma for a $|V_{bias}|>\SI{18}{\kilo\volt}$ and for different pressures. This experimentally confirms that, without a potential well there is no plasma generation in T-REX, and that no voltage breakdown happen based on Paschen's law.
$I_{O.E.}$ is below $\SI{2}{\milli\ampere}$ for all cases, and tends to highlight a growing trend along with $V_{bias}$. Larger amplitudes are achieved when lowering the magnetic field. $I_{T.F.}$ has the largest contribution and grows along with $V_{bias}$ up to $I_{T.F.}<\SI{12}{\milli\ampere}$. $I_{F}$, instead, is zero for all cases. This is due to the topology of the electric field created by the electrodes that, along the direction of the magnetic field, is negative. Therefore, the electrons are accelerated towards the top flange and cannot reach the current probe, that is placed downwards. $I_{Sum}$ is plotted to highlight the small systematic error, ensuring the current reconstruction to be correct for the given accuracy.

Summarizing, the measurements highlight the following:
\begin{enumerate}
    \item $I_{TREK}, I_{O.E.}, I_{T.F.}$ increase with increasing $V_{bias}$;
    \item For the same $V_{bias}$, $I_{TREK}, I_{T.F.}$, and $I_{O.E.}$ increase for decreasing magnetic fields;
    \item $I_{F}$ is zero for all cases;
    \item For $B\leq 0.05 - \SI{0.10}{\tesla}$, or ${|V_{bias}|}\leq 1.5-\SI{2}{\kilo\volt}$, there is no electron cloud and all the currents are zero.
\end{enumerate}

The behavior of point 1. is expected, as the power is increased when increasing the voltage, and it is a trend also observed in FENNECS simulations. However, point 2. is in contradiction with what is predicted by FENNECS simulations: the code predicts that the current should increase with increasing magnetic fields. Diocotron instabilities are a good candidate to describe such divergence in the results, as they could promote the loss of the electron cloud and become even more active at larger values of $B$.Those are instabilities with an azimuthal mode structure that can arise in non-neutral plasmas due to shear, in this case, in the electron cloud angular velocity~\cite{guilth,Eggleston,Danielson}. However, FENNECS cannot model diocotron instabilities as it is a 2D $r,z$ code, therefore we are working on both simulation and experimental side to extend the model including diocotron instabilities. Point 3, according to FENNECS simulations, very few electrons escape downwards from the cloud and reach the current probe due to the electric field shape as mentioned earlier, therefore the results are in line with the simulations. Point 4, the minimum magnetic field for the electron cloud to form under those conditions is of $B\sim\SI{0.10}{\tesla}$ according to FENNECS results, which is in the same order of magnitude of what experiments show. For the electric field, a minimum voltage in the order of $|V_{bias}|>\SI{1.5}{\kilo\volt}$ is required according to FENNECS for the cloud to form, the experiment requires a similar $|V_{bias}|>1.5 - \SI{2}{\kilo\volt}$ to produce the electron cloud.

\section{Comparison Between Experiments and Simulations}

Hereby, we present some preliminary comparison between experiments and simulation via FENNECS. The brevity of this section is due to the nature of the article being a description of T-REX and its first experimental results. Further quantitative comparison with FENNECS is planned for future publications. 

\begin{figure}
\centering
    \begin{subfigure}{0.45\textwidth}
    \includegraphics[width=1.1\linewidth]{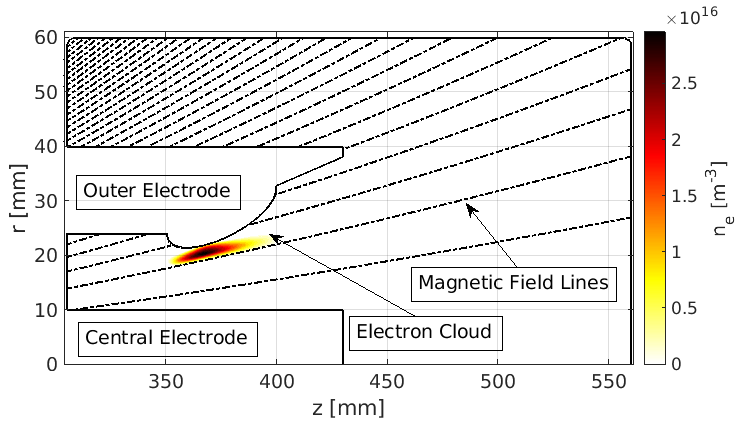}
    \caption{}
     \label{fig:FENNECS_res1}
    \end{subfigure}
     \begin{subfigure}{0.45\textwidth}
    \includegraphics[width=.8\linewidth]{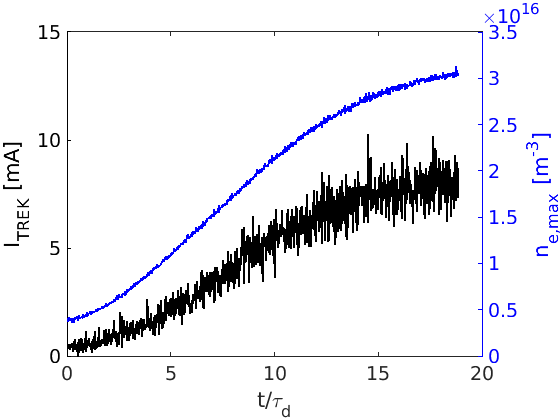}
    \caption{}
    \label{fig:FENNECS_res2}
    \end{subfigure}
    \caption{FENNECS simulation results for T-REX: $V_{bias}=-\SI{8}{\kilo\volt}$ for $B=\SI{0.10}{\tesla}$ operating on He at $p=\SI{5e-4}{\hecto\pascal}$. (a) Simulation results of the electron cloud density $n_e$ in 2D $(r,z)$. (b) Simulation results of the time evolution of electron cloud maximum density $n_{e,max}$ and TREK current $I_{TREK}$.}
    \label{fig:FENNECS_res}
\end{figure}

Fig.~\ref{fig:FENNECS_res1} displays the expected electron cloud density and geometry of T-REX in steady state. In Fig.~\ref{fig:FENNECS_res2}, the simulated $I_{TREK}$ and maximum cloud electron density $n_{e,max}$ are plotted over time during the cloud formation time. Both figures are for T-REX operating on He at $V_{bias}=-\SI{8}{\kilo\volt}$ for $B=\SI{0.10}{\tesla}$. 
The time is represented normalized with respect to the collision characteristic timescale for the momentum exchange $\tau_d=\SI{1.6e-6}{\second}$ in this case. The steady state is defined as the time at which the charge inside the electron cloud saturates and the cloud current oscillates slightly $<10\%$, corresponding to about $20\times \tau_d$~\cite{guilth}.

\begin{figure}
    \centering
    \includegraphics[width=1\linewidth]{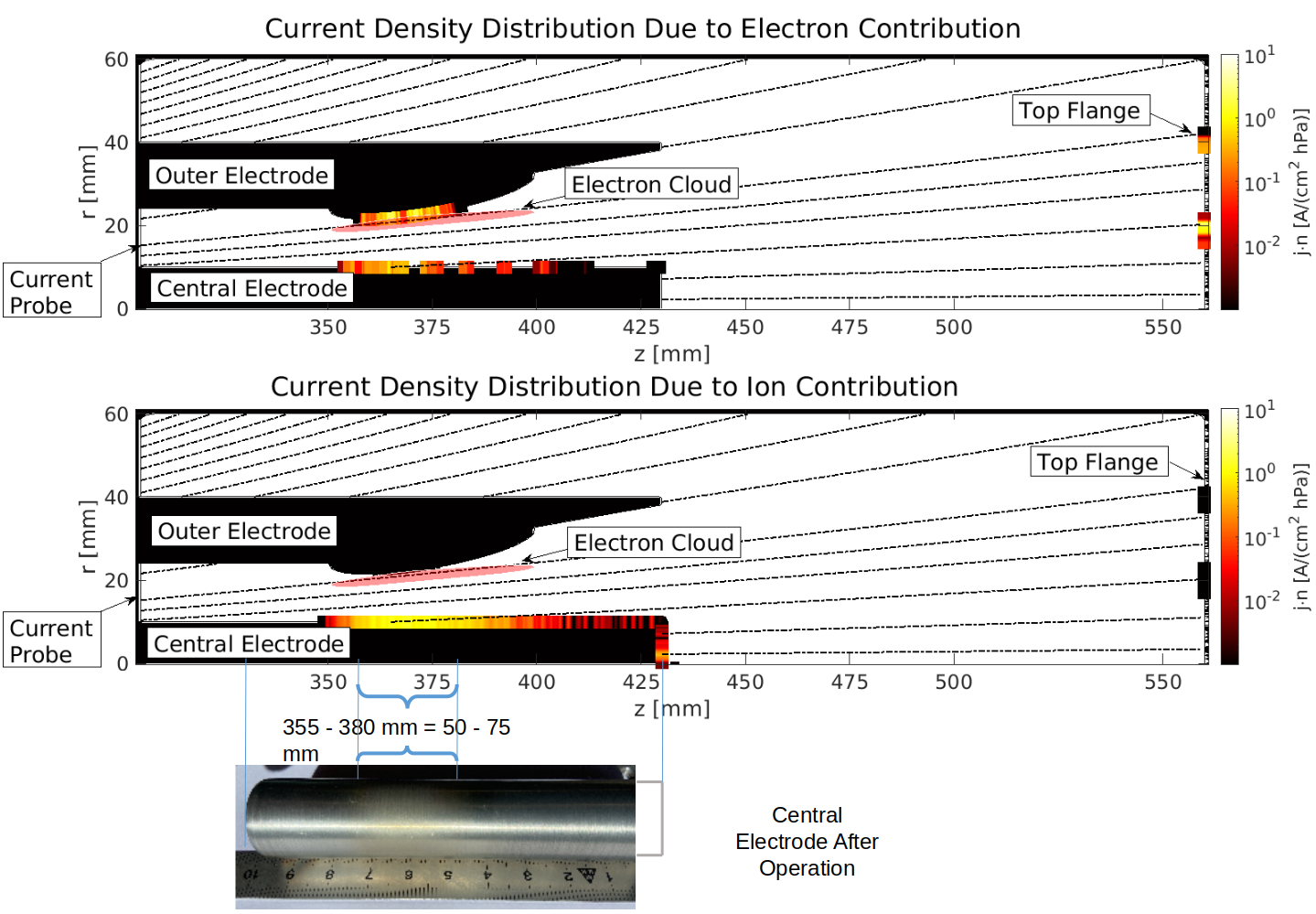}
    \caption{FENNECS simulations of T-REX: current density distribution due to electrons (top) and ions (bottom) contribution. $V_{bias}=-\SI{8}{\kilo\volt}$, $B=\SI{0.10}{\tesla}$, helium at $p=\SI{3E-4}{\hecto\pascal}$. The electron cloud is represented in red, electrons leaving the cloud and those emitted due to ion bombardment follow the magnetic field lines and are then found on the top flange. At the bottom, the effect of ion bombardment is highlighted on the central electrode after operation.}
    \label{fig:electron_currents}
\end{figure}

FENNECS can predict the formation of the electron cloud in steady state under the condition at which T-REX is tested. Furthermore, the order of magnitude of $I_{TREK}$ predicted by FENNECS matches with the experimental test results with T-REX.
In Fig.~\ref{fig:electron_currents}, the current distribution due to electron contribution resulting from FENNECS is shown. The simulations conditions are of T-REX operating on He at $V_{bias}=-\SI{8}{\kilo\volt}$ for $B=\SI{0.10}{\tesla}$. On the central electrode, the current is due to the electrons emitted by ion induced electron emission (IIEE), on the outer electrode it is the current resulting from the electron cloud. On the top flange, the current distribution due to the electrons escaping the electron cloud region following the magnetic field lines can be seen. The largest contribution is the one at the top flange due to IIEE at the central electrode. Also, a current distribution can be seen on the top of the central electrode, while no current is collected at the current probe place at the bottom of the electrodes assembly. Such results are in line with what is measured in the T-REX experiment, namely a large total contribution of current, relative to $I_{TREK}$, on the top flange $I_{T.F.}$ and no current at the current probe $I_F$. Furthermore, a ring of some deposit is seen on the top of the central electrode as well, see Fig.~\ref{fig:T-REXin} previously, that can match with the current distribution on the top of the central electrode resulting from FENNECS.
During the commissioning phase of T-REX it has been necessary to disassemble the electrode assembly multiple times. A change of colour on the aluminium central electrode has been observed in the region where the electron cloud is supposed to form. Indeed, simulations have been run with FENNECS to visualize the current density distribution due to the ions contribution, since the central electrode is biased negatively. The ions are those created by the electrons in the cloud that ionize the residual gas. The results highlights the location, along the axis of symmetry on the central electrode, where the current density is important. This region corresponds qualitatively to the one where the central electrode highlights a different, yellowish colour. The effect is probably due to ion bombardment, which mechanism is included in FENNECS, and can be seen as a further qualitative verification of FENNECS against experiment, see Fig.~\ref{fig:electron_currents}. 

\begin{figure}
    \centering
    \includegraphics[width=.7\linewidth]{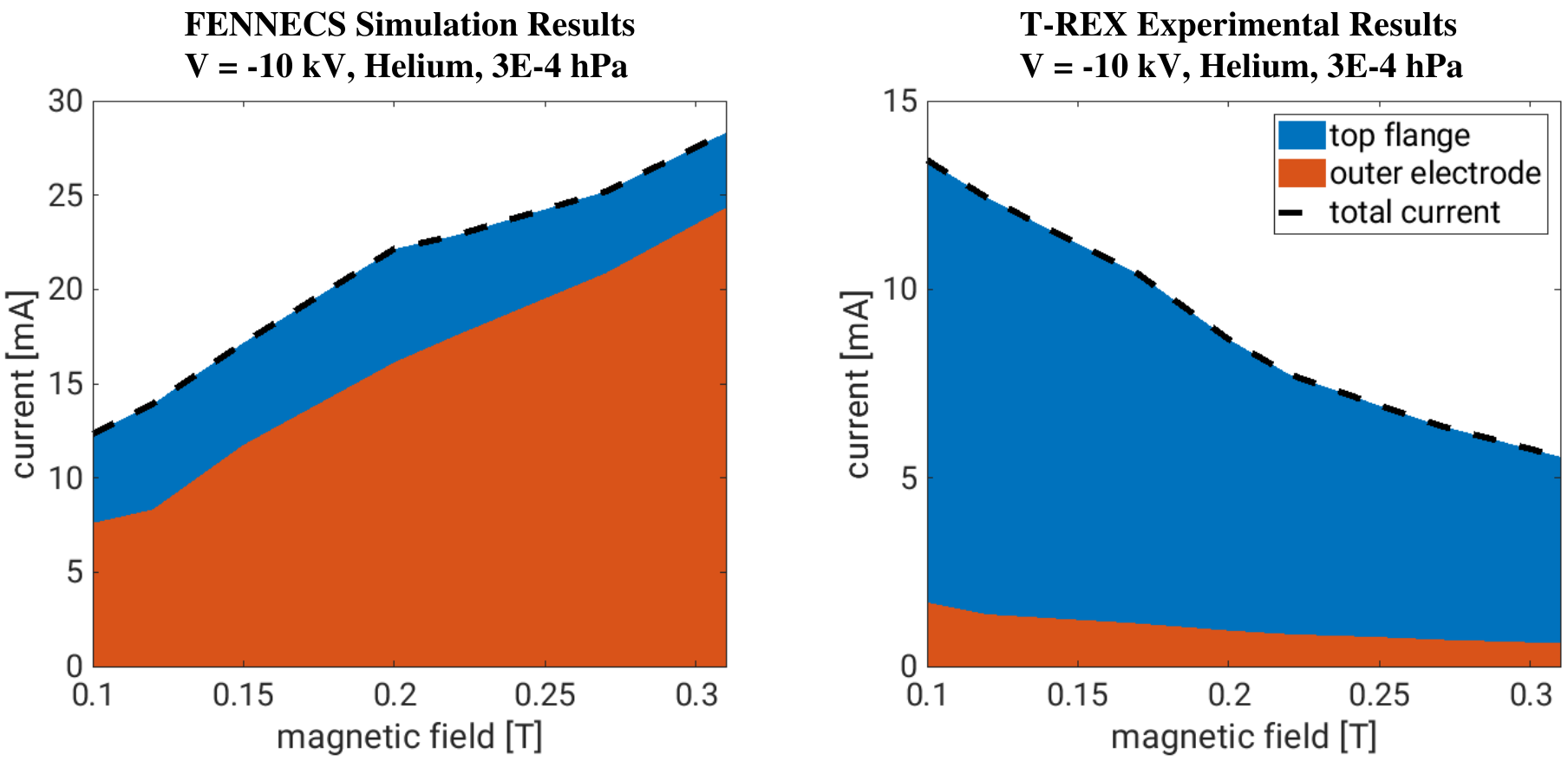}
    \caption{Plots showing the trend of top flange, outer electrode, and total currents for different magnetic fields at $V_{bias}=-\SI{10}{\kilo\volt}$ on He at $p=\SI{3E-4}{\hecto\pascal}$. On the left the simulation results, on the right the experimental results.}
    \label{fig:T-REX_B}
\end{figure}

Also, T-REX highlights diminishing currents for larger magnetic fields, while FENNECS predicts the opposite, this can be seen in Fig.~\ref{fig:T-REX_B} that shows the trend of currents simulated via FENNECS vs those measured experimentally in T-REX. This is still currently under investigation and requires further studies. As mentioned earlier, our first working hypothesis points toward  diocotron instabilities and we are working on a model to simulate them.

\section{Future Diagnostics}

\begin{figure}
    \centering
    \includegraphics[width=.8\linewidth]{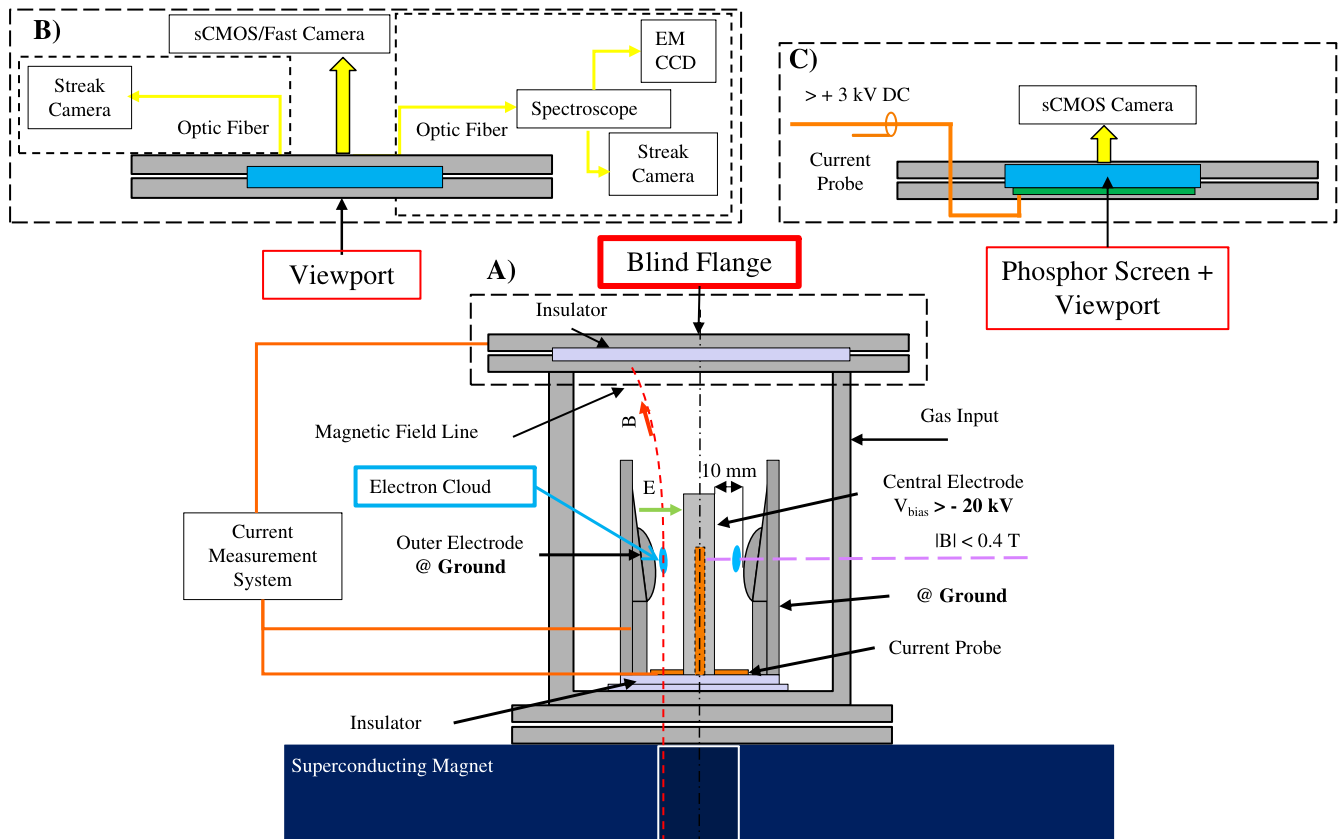}
    \caption{T-REX set-up and its diagnostics schematics. \textbf{In use: A)}: Blind stainless steel flange: currents and voltage measurements. \textbf{Planned: B)}: Optical diagnostics via sCMOS, fast, or streak camera; OES via EM CCD or streak camera. \textbf{Planned: C)} Phosphor screen total charge measurement and imaging via sCMOS.}
    \label{fig:T-REX_SCHEMATICS}
\end{figure}

The current measurement system is still under optimization in terms of reliability and faster acquisition, with a target cut-off frequency $>\SI{300}{\kilo\hertz}$ to capture faster cloud dynamics. Another objective is of measuring $I_{TREK}$ independently with a higher cut-off frequency to capture time behavior as, currently, only the DC component can be seen through the TREK amplifier. In addition, a magnetic-field-independent pressure measurement installed as close as possible to the cloud region is being designed to better visualize how the pressure changes at different experimental conditions. An array of Langmuir probes will be installed on top of the chamber to measure the radial current/energy distribution of the electrons that streams upward from the electron cloud. Concerning the optical diagnostics, it is planned to install diagnostics such as scientific Complementary Metal–Oxide–Semiconductor (sCMOS) and fast camera, in the $\SI{}{\micro\second}$ range time resolution, to perform imaging of the electron cloud from the top to evaluate its size and thickness at different experimental conditions and its dynamics in combination with the current measurements. Optical emission spectroscopy (OES) via electron-multiplying charge-coupled device (EM CCD) is planned to identify the species and ionization states and, when looked through a streak camera, evaluate the spectra over time behaviour. The streak camera imaging as shown in Fig.~\ref{fig:T-REX_SCHEMATICS}, \textbf{B)} will be used to visualize temporal variation of size and brightness of the electron cloud. Those can be foreseen in terms of the plasma characteristic frequencies that, given FENNECS results, can be in the range of $1-\SI{11}{\giga\hertz}$ (bounce to cyclotron frequencies). OES will also be used to determine the radial electric field distribution via the Stark effect, in particular by using Hydrogen as background gas, the $H-\alpha$ line could have a maximum extension of its emission's wavelengths $\lambda= 655.5 - \SI{656.8}{\nano\meter}$ at $V_{bias}=-\SI{20}{\kilo\volt}$ and $B=\SI{0.1}{\tesla}$.

A biased phosphor screen based diagnostics will also be installed, Fig.~\ref{fig:T-REX_SCHEMATICS} \textbf{C)}. It will be used to evaluate the total charge and to extract eventual structures by dumping the cloud on the screen itself. The removal of the electric field, will be performed via a \SI{200}{\nano\second} HV switch. The experiment is expected to react slower due to stray capacities. This is measured in the \SI{}{\micro\second} range, and evaluation via faster current measurement will be performed to assess how this compares with the electron cloud dynamics timescale. The image of the dumped electron cloud on the phosphor screen will also be used to determine its size by also knowing the divergence of the magnetic field.
The planned optical diagnostics can be resumed as following:
\begin{itemize}
    \item A viewport is installed on top of the vacuum chamber: imaging of the electron cloud via a sCMOS camera, fast camera, or via the streak camera, see \textbf{B)} in Fig.~\ref{fig:T-REX_SCHEMATICS};
    \item OES via EM CCD or via the streak camera \textbf{B)} in Fig.~\ref{fig:T-REX_SCHEMATICS};
    \item A biased ($V>\SI{3}{\kilo\volt}$) phosphor screen mounted in front of the viewport on top of the vacuum chamber: imaging via sCMOS camera after dumping the electron cloud on the screen by removing $V_{bias}$ at the central electrode (therefore removing the confinement) \textbf{C)} in Fig.~\ref{fig:T-REX_SCHEMATICS}.
\end{itemize}

On the simulation side, FENNECS will be improved via further validation against experiments and added physics.

\section{Conclusion}
A novel and unique non-neutral plasma physics experiment, the TRapped Electrons eXperiment (T-REX), that aims at reproducing typical gyrotron MIG conditions to study trapped secondary electrons has been simulated, designed, built, integrated, and tested. The experiment has been successfully operated and produces a trapped electron plasma in the desired region for multiple voltages, gases, magnetic field strengths and pressures. The first successful measurements of the current distribution within the experiment main components have been done. Results show agreeing trends between test conditions, in most of the current is collected on the top flange of the vacuum chamber. 

Furthermore, we observe a solid agreements between the FENNECS code and the T-REX experiment in the following:
\begin{itemize}
    \item the formation of the electron cloud with the given T-REX configuration;
    \item the order of magnitude ($1-\SI{20}{\milli\ampere}$) of the measured currents; 
    \item the trend of growing total input current from the power supply $I_{TREK}$ along with the applied voltage $V_{bias}$;
    \item the extension (and location) of the area along the central conductor where (possible) ion bombardment is higher;
    \item the qualitative current distribution due to electrons; 
    \item the minimum magnetic field $B>\SI{0.1}{\tesla}$ and minimum bias$|V_{bias}|>\SI{1.5}{\kilo\volt}$  required for the electron cloud to form.
\end{itemize}
From the results of T-REX experiments, the cloud formation mechanisms seem to be consistent with the picture emerging from FENNECS (this is visible from the position of the cloud and the formation thresholds in terms of minimum $B$ and $V_{bias}$), but the loss mechanisms seem to require additional physics that are not captured by FENNECS at this moment (this is visible from the fact that more electrons are lost on the top flange than on the outer electrode). Continuous improvement of the model is foreseen in the near future.

\begin{acknowledgments}
This work was supported by the Swiss National Science Foundation under grant No.~204631.\\

This work has been carried out within the framework of the EUROfusion Consortium, via the Euratom Research and Training Programme (Grant Agreement No.~101052200 - EUROfusion) and funded by the Swiss State Secretariat for Education, Research and Innovation (SERI). Views and opinions expressed are, however, those of the author(s) only and do not necessarily reflect those of the European Union, the European Commission, or SERI. Neither the European Union nor the European Commission nor SERI can be held responsible for them.\\

The article has been accepted by Review of Scientific Instruments. 

\end{acknowledgments}

\appendix
\section{Estimation of the Recombination and Excitation/De-Excitation Rates}
In this section, the recombination $\nu_{Rec}$ and excitation/de-excitation $\nu_{Exc}$ rates are estimated, see Eq.~\ref{eq:nurec} and Eq.~\ref{eq:nuexc} respectively, to determine the dominant mechanism behind plasma light emission in T-REX. As anticipated earlier in the text, it is expected that recombination is negligible.\\

The hypotheses are the following:
\begin{itemize}
    \item The electric field is the order of $|\vec{E}|\sim V_{bias}/d\sim\SI{1}{\mega\volt\per{\meter}}$ taking $V_{bias}\sim\SI{10}{\kilo\volt}$ and $d\sim\SI{1}{\centi\meter}$, with $d$ the distance between the two electrodes; the magnetic field $\vec{B}$ is constant and uniform at $B=\SI{0.17}{\tesla}$. This gives a drift velocity in the order of $v_e = E/B \sim \SI{5.88E6}{\meter\per{\second}}$;
    \item The cross sections are evaluated for Helium for a kinetic energy of $E_k \sim \frac{1}{2} \frac{m_e v_e^2}{q_e} = \SI{100}{\electronvolt}$ with $m_e$ and $q_e$ the mass and the charge of electrons:
    \begin{itemize}
        \item $\sigma_{Exc}\approx\SI{7.3E-23}{}-\SI{1.8E-21}{\meter^2}$, excitation cross section~\cite{cross_sections};
        \item $\sigma_{Rec}\approx\SI{1E-27}{\meter^2}$, radiative recombination cross section~\cite{janev2012elementary};
        \item $\sigma_{Ion}\approx\SI{3.6E-21}{\meter^2}$, ionization cross section~\cite{cross_sections};
    \end{itemize}
    \item The electron cloud has a constant density in the order of $n_e\sim\SI{1E18}{\meter^{-3}}$;
    \item An ion created by ionization has 0 initial velocity and travels to the central electrode due to the $\vec{E}$-field only. The travel time for one ion to the central electrode $\tau_{Loss,ion}$ is calculated as $\tau_{Loss,ion} = \sqrt{{2 d m_{He,ion}}/{q_e E}}$;
    \item The neutral pressure is constant and uniform at $p=\SI{3E-4}{\hecto\pascal}$, the neutral density $n_n$ is then estimated via the ideal gas law $n_n = p / (k_B T)$, with $T\approx\SI{300}{\kelvin}$.  
\end{itemize}

Calculation of the recombination rate $\nu_{Rec}$:
\begin{equation}
    \nu_{Rec} = n_i \sigma_{Rec} v_e 
    \label{eq:nurec}
\end{equation}
Calculation of the excitation/de-excitation rate $\nu_{Exc}$:
\begin{equation}
    \nu_{Exc} = n_e \sigma_{Exc} v_e 
    \label{eq:nuexc}
\end{equation}
The ion density $n_i$, see Eq.~\ref{eq:ni}, can be estimated as the product of $R = n_e n_n \sigma_{Ion} v_e$, the number of ions created by ionization, per second and per $\SI{}{\meter^3}$, multiplied by the time it takes for such ions to be lost at the central electrode $\tau_{Loss,ion}$. Thus: 
\begin{equation}
    n_i = R \tau_{Loss,ion}
    \label{eq:ni}
\end{equation}
For the given values, the recombination and excitation/de-excitation rates can be calculated:
\begin{equation}
    \nu_{Exc} = \SI{3.1E3}{} - \SI{7.5E4}{\per\second}
\end{equation}
\begin{equation}
    \nu_{Rec} = \SI{2.6E-5}{\per\second}
\end{equation}
Therefore,
\begin{equation}
    \nu_{Exc}>>\nu_{Rec}
\end{equation}
This numerical estimation indicates that excitation and de-excitation processes are predominant compared to the contribution of recombination events. Consequently, it can be said that the primary source of light emission arises from the excitation-de-excitation transitions rather than from recombination processes.

\nocite{*}
\section*{References}
\bibliography{aipsamp}

\end{document}